\documentclass[letterpaper]{article}

\usepackage[T1]{fontenc}

\usepackage{geometry}
\usepackage{setspace}

\usepackage{graphicx}

\usepackage[articletitle = true,
            chaptertitle = false,
            doi          = false,
            maxauthors   = 15,
            etalmode     = firstonly,
            biblabel     = brackets,
            super        = true]{achemso}

\usepackage[version=3]{mhchem} %
\usepackage{siunitx}
\usepackage{url}%
\usepackage{xspace}
\DeclareSIUnit{\atom}{atom}
\DeclareSIUnit{\angstrom}{\text{\AA}}

\usepackage{authblk}
\usepackage{pdfpages}

\usepackage{cleveref}
\crefname{figure}{Figure}{Figures}
\crefname{table}{Table}{Tables}
\crefformat{equation}{Eq.~#2#1#3}
\Crefformat{equation}{Eq.~#2#1#3}

\newcommand*{\rtwoscan}{r\textsuperscript{2}SCAN\xspace}
\newcommand*{\gnAlOCNmax}{5}%
\newcommand*{\gnAlOCNmin}{3}%
\newcommand*{\gnAlOSigmaEbMax}{13.4}%
\newcommand*{\gnAlOSigmaEbMin}{8.1}%
\newcommand*{\gnCbRT}{+0.049}%
\newcommand*{\gnCbRTStd}{0.040}%
\newcommand*{\gnCgRT}{+0.23}%
\newcommand*{\gnChullAll}{+2.9}%
\newcommand*{\gnChullAllStd}{4.3}%
\newcommand*{\gnChullRT}{+6.2}%
\newcommand*{\gnChullRTStd}{6.0}%
\newcommand*{\gnCrMinusParent}{0.2}%
\newcommand*{\gnCrNative}{-1.67}%
\newcommand*{\gnDminAddDMae}{-0.024}%
\newcommand*{\gnDminR}{+0.65}%
\newcommand*{\gnDminRLooMax}{+0.67}%
\newcommand*{\gnDminRLooMin}{+0.60}%
\newcommand*{\gnDropBothAllMean}{-0.011}%
\newcommand*{\gnDropBothAllPctLower}{64}%
\newcommand*{\gnDropBothRTMean}{-0.014}%
\newcommand*{\gnDropBothRTPctLower}{67}%
\newcommand*{\gnDropEbRTMean}{-0.006}%
\newcommand*{\gnDropEbRTPctLower}{52}%
\newcommand*{\gnDropEgRTMean}{+0.055}%
\newcommand*{\gnDropEgRTPctLower}{7}%
\newcommand*{\gnEgMax}{8.5}%
\newcommand*{\gnEhullMPMaxDiffRT}{6}%
\newcommand*{\gnEhullRTmax}{23}%
\newcommand*{\gnEhullRTmin}{0}%
\newcommand*{\gnEvMeanGroupFifteen}{2.8}%
\newcommand*{\gnEvMeanGroupFourteen}{2.7}%
\newcommand*{\gnEvMeanGroupOne}{5.2}%
\newcommand*{\gnEvMeanGroupTwo}{7.2}%
\newcommand*{\gnEvPbPbcn}{0.74}%
\newcommand*{\gnEvPbRut}{0.09}%
\newcommand*{\gnEvSpan}{8}%
\newcommand*{\gnEvTlIa}{0.48}%
\newcommand*{\gnExpRouteDiffMax}{0.07}%
\newcommand*{\gnExpRouteDiffMin}{0.04}%
\newcommand*{\gnGaOSpread}{0.9}%
\newcommand*{\gnLadderARTUnc}{0.42(7)}%
\newcommand*{\gnLadderERTUnc}{0.33(5)}%
\newcommand*{\gnLocoAllPooled}{0.29}%
\newcommand*{\gnLocoRTPooled}{0.31}%
\newcommand*{\gnLohoDminAll}{39}%
\newcommand*{\gnLohoDminGapAll}{44}%
\newcommand*{\gnLopoAllPooled}{0.28}%
\newcommand*{\gnLopoRTPooled}{0.30}%
\newcommand*{\gnMaeFourAllUnc}{0.30(3)}%
\newcommand*{\gnMaeFourRTUnc}{0.33(4)}%
\newcommand*{\gnMaeMetastable}{0.21}%
\newcommand*{\gnMaePrimAllUnc}{0.28(3)}%
\newcommand*{\gnMaePrimRTUnc}{0.31(4)}%
\newcommand*{\gnMaePrimary}{0.31}%
\newcommand*{\gnMaeVrOnlyRTUnc}{0.37(5)}%
\newcommand*{\gnMetaGroundStateComps}{\ce{SnO}}%
\newcommand*{\gnMetallicFanSpread}{6}%
\newcommand*{\gnMetallicLowestPhrase}{the three lowest}%
\newcommand*{\gnMetallicOverMax}{1.3}%
\newcommand*{\gnMetallicOverMin}{0.6}%
\newcommand*{\gnNAtomsMax}{240}%
\newcommand*{\gnNAtomsMin}{80}%
\newcommand*{\gnNCenHosts}{14}%
\newcommand*{\gnNCenPts}{47}%
\newcommand*{\gnNCompRetained}{24}%
\newcommand*{\gnNExclDuplicateWord}{two}%
\newcommand*{\gnNExclHighPressureWord}{ten}%
\newcommand*{\gnNExclIdealizedWord}{nine}%
\newcommand*{\gnNExclNonBulkWord}{one}%
\newcommand*{\gnNExclNotBinaryWord}{five}%
\newcommand*{\gnNHostsThreeWord}{nine}%
\newcommand*{\gnNIcsdListed}{467}%
\newcommand*{\gnNPolyClean}{39}%
\newcommand*{\gnNPolyExcluded}{27}%
\newcommand*{\gnNPolyRetained}{39}%
\newcommand*{\gnNPolyTotal}{66}%
\newcommand*{\gnNRefcheckHosts}{39}%
\newcommand*{\gnNSLike}{22}%
\newcommand*{\gnNShallowWord}{four}%
\newcommand*{\gnNSingleBand}{70}%
\newcommand*{\gnNSitesBiSevenWord}{seven}%
\newcommand*{\gnNSitesClean}{72}%
\newcommand*{\gnNSitesEstimatedDHf}{13}%
\newcommand*{\gnNSitesGaMonoWord}{three}%
\newcommand*{\gnNSitesMeta}{16}%
\newcommand*{\gnNSitesPbMonoWord}{three}%
\newcommand*{\gnNSitesRT}{56}%
\newcommand*{\gnNSymChangeWord}{ten}%
\newcommand*{\gnParentChullAll}{-5.8}%
\newcommand*{\gnParentChullRT}{-55.8}%
\newcommand*{\gnParentCr}{-1.5}%
\newcommand*{\gnParentMaeAll}{0.70}%
\newcommand*{\gnParentMaeAllStd}{0.04}%
\newcommand*{\gnParentMaeRT}{0.45}%
\newcommand*{\gnParentMaeRTStd}{0.04}%
\newcommand*{\gnPartialCNGivenDmin}{+0.05}%
\newcommand*{\gnPartialDminGivenCN}{+0.57}%
\newcommand*{\gnPrimCg}{0.25}%
\newcommand*{\gnPrimCr}{-1.74}%
\newcommand*{\gnPrimCzero}{0.91}%
\newcommand*{\gnREbDFT}{+0.76}%
\newcommand*{\gnREgDFT}{+0.88}%
\newcommand*{\gnREhullDFT}{+0.30}%
\newcommand*{\gnRRedVrAll}{-0.999}%
\newcommand*{\gnRTwoFourAll}{0.97}%
\newcommand*{\gnRTwoFourRT}{0.97}%
\newcommand*{\gnRTwoPrimAll}{0.97}%
\newcommand*{\gnRTwoPrimRT}{0.97}%
\newcommand*{\gnRVrDFT}{-0.98}%
\newcommand*{\gnResidRmsRT}{0.18}%
\newcommand*{\gnSigmaEb}{2.04}%
\newcommand*{\gnSigmaEg}{1.85}%
\newcommand*{\gnSigmaEhull}{7.9}%
\newcommand*{\gnSigmaVr}{0.96}%
\newcommand*{\gnSiteVrMaeRT}{0.48}%
\newcommand*{\gnSiteVrMaeRTBase}{0.33}%
\newcommand*{\gnSnOAllenDiff}{0.3}%
\newcommand*{\gnSnOAllenVO}{4.0}%
\newcommand*{\gnSnOLithEv}{4.29}%
\newcommand*{\gnSnOLithPred}{3.40}%
\newcommand*{\gnSnOLithUnder}{0.9}%
\newcommand*{\gnSnOtwoMinusSnO}{0.3}%
\newcommand*{\gnSpanBiSeven}{0.89}%
\newcommand*{\gnSpanGaMono}{0.77}%
\newcommand*{\gnSpanPbMono}{0.97}%
\newcommand*{\gnStdWeightVr}{-1.60}%
\newcommand*{\gnSweepPrimAll}{0.29}%
\newcommand*{\gnSweepPrimFifty}{0.28}%
\newcommand*{\gnSweepPrimHundred}{0.28}%
\newcommand*{\gnSweepPrimTwentyFive}{0.31}%
\newcommand*{\gnTlOThreeDimerDE}{7.5}%

\author[1]{Manish Kumar}
\author[1]{Bridget S.\ Ferris}
\author[1,2]{Avital Isakov}
\author[1]{Robert B. Wexler*}
\affil[1]{Department of Chemistry and Institute of Materials Science and Engineering, Washington University in St.~Louis, St.~Louis, MO 63130, USA}
\affil[2]{Present address: Department of Chemistry and Biochemistry, University of California, Los Angeles, Los Angeles, CA 90095, USA}

\title{Periodic Trends in Oxygen Vacancy Formation Energies Across Binary Main-Group Metal Oxides}
\date{*Email: wexler@wustl.edu}
\begin{document}

\maketitle

\begin{abstract}
  The function of oxide materials, from solid-oxide fuel cells to solar thermochemical fuel production, depends on their oxygen vacancies.
  Their formation energetics have been mapped systematically for ternary perovskites, but reported values for the binary oxides, their conceptual and thermochemical building blocks, remain scattered across studies with differing computational methods.
  Here we report a systematic \rtwoscan{} density functional theory survey of neutral oxygen vacancy formation energies across \num{\gnNPolyRetained} binary main-group metal oxide polymorphs spanning groups 1, 2, 13, 14, and 15, using the \textsc{ShakeNBreak} method to locate reconstructed defect ground states.
  The formation energies span \qty{\gnEvSpan}{\eV} and decrease down each group and, for most multivalent cations, with increasing oxidation state.
  Both trends follow the reducibility of the cation.
  An interpretable crystal feature model is fit to these energies; its inputs are properties of the pristine bulk, with the reduction potential alternatively taken from tabulated thermochemistry.
  Retaining only its two dominant descriptors, the crystal reduction potential and the band gap, the model reproduces the formation energies with a held-out mean absolute error of \qty{\gnMaePrimary}{\eV}, lower than that of its four-descriptor parent model for transition-metal perovskites.
  The bond-breaking and stability terms of the full Born--Haber fit prove statistically null for the binary oxides, and the largest residuals identify the hosts where electron accommodation departs from the cation-reduction picture that the descriptors encode.
  The dataset provides a benchmark for machine-learning interatomic potentials, and the model estimates vacancy formation energies of main-group oxides from bulk properties alone.
\end{abstract}

\section{Introduction}

Oxygen vacancies in metal oxides mediate performance across a broad range of energy technologies.
In solid oxide fuel cells, vacancies provide pathways for bulk oxide ion transport and determine the rate of surface oxygen exchange,\cite{deml_oxide_2014, wexler_factors_2021} and in solar thermochemical cycles, the vacancy formation energy determines the operating temperatures and thermodynamic efficiency of hydrogen production.\cite{emery_highthroughput_2016, wexler_factors_2021, baldassarri_oxygen_2023}
Oxygen vacancies are also implicated in the $n$-type conductivity of transparent conducting oxides, where their deep versus shallow character has been debated,\cite{linderalv_unifying_2018} in the formation of catalytically active sites for oxidation reactions,\cite{tukur_machine_2025} and in the degradation of lithium-ion battery cathodes through oxygen loss.\cite{baldassarri_oxygen_2023}
A systematic mapping of vacancy formation energetics across oxide chemical space nevertheless remains incomplete.

The energy required to form a neutral oxygen vacancy can be understood through an extended Born--Haber cycle involving two principal contributions.\cite{deml_oxide_2014, wexler_factors_2021}
First, the metal--oxygen bonds coordinating the vacancy site must be broken, an energetic cost that scales with the strength of local M--O bonding.
Second, the two electrons formerly associated with the lattice \ce{O^2-} ion must be accommodated elsewhere in the structure, typically by reducing neighboring cations or, in metallic systems, by populating states near the Fermi level.
The interplay between these bond-breaking and electron-accommodation terms explains why vacancy formation energies correlate with both thermodynamic quantities such as the oxide formation enthalpy and electronic structure features including the band gap and the position of the oxygen 2p band.\cite{deml_oxide_2014, deml_intrinsic_2015}
Building on this physical picture, Wexler et al.\cite{wexler_factors_2021} decomposed the formation energy into contributions from crystal bond dissociation energies, crystal reduction potentials, band gaps, and thermodynamic stability, achieving mean absolute errors of approximately \qty{0.45}{\eV} for perovskite oxides stable at ambient conditions.
Linder\"alv et al.\cite{linderalv_unifying_2018} demonstrated that the charge transition levels of oxygen vacancies, when aligned to the vacuum level, exhibit small variation across wide-band-gap binary and ternary oxides, pointing to underlying universality in the electronic structure of the defect state itself.
These conceptual advances notwithstanding, translating physical insight into broadly predictive models has proven challenging, in part because the relative importance of local descriptors (properties of the immediate coordination environment of the vacant oxygen) and global descriptors (properties of the host crystal as a whole) varies with crystal structure and composition.

High-throughput computational studies have mapped vacancy formation energies across increasingly large regions of oxide chemical space.
Emery and Wolverton\cite{emery_highthroughput_2017} computed formation energies for \num{5329} \ce{ABO3} perovskites.
Kumagai et al.\cite{kumagai_insights_2021} performed automated point-defect calculations on \num{937} oxides (approximately \num{6000} vacancy calculations, three charge states at each symmetry-inequivalent oxygen site) and trained random forest models with mean absolute errors of \qtyrange[range-phrase = --, range-units = single]{0.27}{0.44}{\eV} depending on charge state, and Baldassarri et al.\cite{baldassarri_oxygen_2023} extended this approach to over \num{2500} vacancy calculations across more than \num{1000} compounds with comparable accuracy.
Graph neural networks have also been applied to vacancy prediction: Witman et al.\cite{witman_defect_2023} developed a defect graph neural network that learns directly from crystal graphs without hand-engineered features, and Choudhary and Sumpter\cite{choudhary_can_2023} demonstrated that a graph neural network energy model trained only on defect-free structures can extrapolate to vacancy formation energies, albeit with reduced accuracy.

Despite this progress, the existing computational literature exhibits a pronounced compositional bias: ternary and quaternary oxides (perovskites in particular) dominate, and the main-group elements of groups 1, 2, 13, 14, and 15 appear sporadically across these datasets but have not been surveyed across all known polymorphs of their binary oxides.
Binary oxides allow structure--property relationships to be examined most directly: with only one cation species, variations in vacancy formation energy across polymorphs can be attributed to differences in local coordination, bond lengths, and electronic structure rather than to compositional complexity.
Prior descriptor-based models have also used the thermochemistry of binary oxides (their formation enthalpies, cation reduction energetics, and vacancy formation energies) as inputs for predicting vacancy behavior in multinary compounds,\cite{wexler_factors_2021, baldassarri_oxygen_2023} and the solid-state electrochemical series of Mueller et al.\cite{mueller_electrochemical_2024} underscores the value of solid-state, rather than aqueous, reference data for organizing redox behavior.
Establishing periodic trends in vacancy formation energies across binary oxide chemical space (analogous to the periodic trends in ionization energies or electron affinities that organize atomic behavior) would provide both physical insight and a foundation for rational defect engineering in more complex systems.

Here we report a systematic computational survey of neutral oxygen vacancy formation energies in binary oxides of the main-group metals spanning groups 1 (\ce{Li}, \ce{Na}, \ce{K}, \ce{Rb}, \ce{Cs}), 2 (\ce{Be}, \ce{Mg}, \ce{Ca}, \ce{Sr}, \ce{Ba}), 13 (\ce{Al}, \ce{Ga}, \ce{In}, \ce{Tl}), 14 (\ce{Sn}, \ce{Pb}), and 15 (\ce{Bi}).
We consider every polymorph of these oxides that the Materials Project\cite{jain_commentary_2013} lists with an experimental structure from the Inorganic Crystal Structure Database (ICSD),\cite{zagorac_recent_2019} apart from one \ce{Bi2O3} entry whose structure contains O--O pairs (see Polymorph Selection), and compute the formation energy of every symmetry-inequivalent oxygen vacancy site using density functional theory with the \rtwoscan{} meta-generalized-gradient functional.
This choice aligns with the current Materials Project computational framework and extends \rtwoscan{} defect calculations, which have so far targeted individual hosts,\cite{abbas_theoretical_2025} to a high-throughput dataset.

We employ the \textsc{ShakeNBreak} approach for defect structure optimization, which samples a diverse set of initial distortions to identify low-energy reconstructed configurations that conventional local relaxation often misses.\cite{mosquera-lois_shakenbreak_2022, mosquera-lois_identifying_2023}
This methodology guards against the systematic overestimation of formation energies that can occur when calculations become trapped in metastable high-symmetry configurations, a source of error that prior high-throughput defect studies, which relaxed each vacancy from a single starting configuration, have not systematically addressed.\cite{kumagai_insights_2021, witman_defect_2023, baldassarri_oxygen_2023}

This work makes three principal contributions.
First, we adapt the interpretable crystal feature model of Wexler et al.\cite{wexler_factors_2021} to the binary oxides. Of its four physically motivated descriptors (the crystal bond dissociation energy, the crystal reduction potential, the band gap, and the energy above the convex hull), the reduction potential and the band gap prove sufficient, and the largest residuals of the model identify hosts in which electron accommodation departs from the cation reduction that the descriptors encode.
Second, the dataset maps vacancy formation energies systematically across the binary oxides of these metal groups and establishes quantitative periodic trends in them: across hosts, the formation energies follow cation reducibility and oxidation state; within a host, they follow the local bond geometry of the vacant site.
Third, we offer this dataset as a benchmark resource for machine-learning interatomic potentials, for which reference data spanning simple binary systems are needed to assess transferability and identify failure modes.

\section{Methods}

\subsection{Polymorph Selection and Training Data Curation}

\begin{figure}
  \centering
  \includegraphics[width=\linewidth]{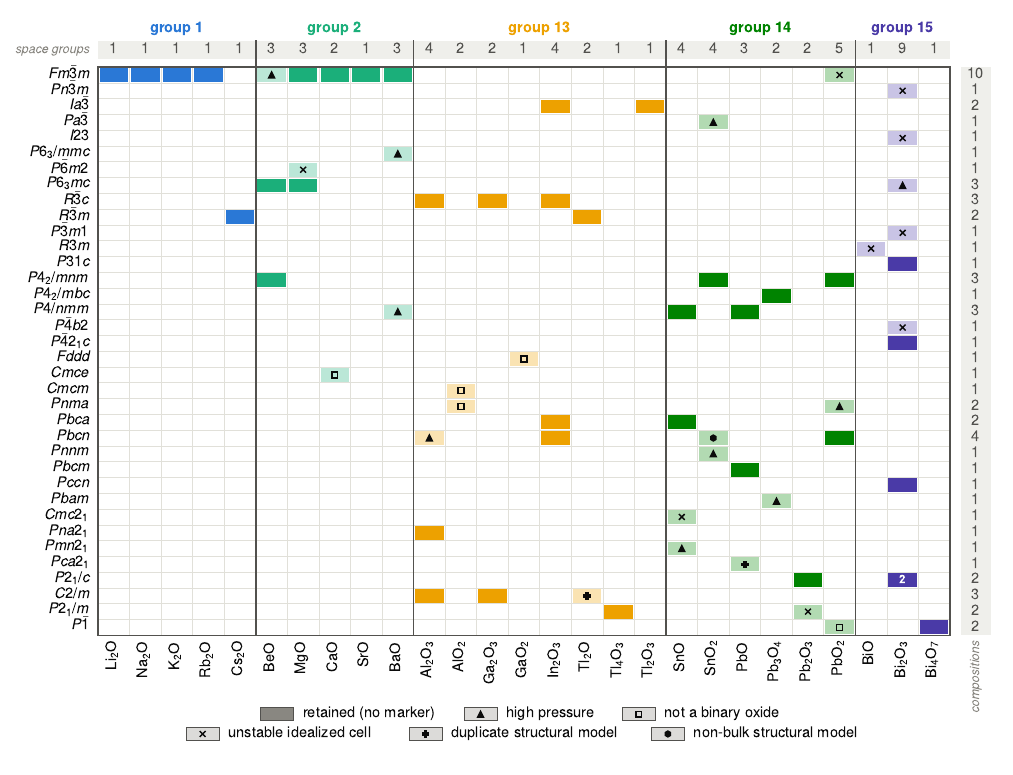}
  \caption{Composition--space-group coverage of the dataset and curation of the training set. Each filled cell holds one of the \num{66} material--polymorph combinations (the cell marked 2 holds two, \(\alpha\)-\ce{Bi2O3} and \ce{Bi2O3}-HP(I), both \(P2_1/c\)), colored by main group (groups 1, 2, 13, 14, and 15); compositions are ordered within each main group by cation atomic number and then by oxygen-to-metal ratio, and space groups are ordered by descending international number. Solid cells mark the \num{\gnNPolyRetained} polymorphs retained for training; tinted cells mark the \num{\gnNPolyExcluded} excluded, with the marker giving the exclusion class as keyed in the legend. Marginal strips count space groups per composition (top) and compositions per space group (right).}
  \label{fig:compsg}
\end{figure}

The dataset comprises \num{66} material--polymorph combinations of \num{27} nominal binary oxide compositions of the \num{17} main-group metals, in all seven crystal systems and \num{36} space groups (\cref{fig:compsg}).
One further \ce{Bi2O3} entry of the Materials Project (mp-558953) was not computed, because it is not a binary oxide of \ce{O^2-}: a third of its oxygen atoms form O--O pairs \qty{1.52}{\angstrom} apart.
Because the crystal feature model (CFM)\cite{wexler_factors_2021} is best suited to hosts that are at least metastable under the near-ambient conditions of typical synthesis and operation, the training set was curated from the experimental provenance of each polymorph in the ICSD.\cite{zagorac_recent_2019}
Every ICSD collection code that the Materials Project lists was retrieved (\num{\gnNIcsdListed} codes in total), and each polymorph was classified according to whether at least one experimentally determined entry corresponds to near-ambient conditions, defined here as a temperature within \qtyrange[range-phrase = --, range-units = single]{273}{313}{\kelvin} and a pressure \qty{\leq 0.2}{\MPa} (classification rules in the Supporting Information).

Of the \num{\gnNPolyTotal} polymorphs evaluated, \num{\gnNPolyRetained} were retained for CFM training, and \num{\gnNPolyExcluded} were excluded, in five classes (\cref{fig:compsg}): \gnNExclHighPressureWord{} phases that exist only under high hydrostatic pressure; \gnNExclNotBinaryWord{} records that are not binary oxides, such as oxyhydroxides recorded without their hydrogen atoms; \gnNExclIdealizedWord{} unstable idealized cells, which are dynamically unstable in the primitive cell or in the vacancy supercell or do not reproduce the recorded structure; \gnNExclDuplicateWord{} earlier structural models of phases retained under another space group; and \gnNExclNonBulkWord{} model of a non-bulk layer.
The high-pressure phases are excluded because the descriptors, like the formation energies, are evaluated at zero pressure, at which these phases are at best metastable; at the pressures where the phases are stable, the vacancy formation enthalpy acquires a pressure--volume term, the product of the pressure and the vacancy relaxation volume, that none of the descriptors represents.
For the records that are not binary oxides and for the unstable idealized cells, no vacancy formation energy of a distinct binary-oxide phase can be defined.
The evidence and the structural references for each excluded polymorph are given in the Supporting Information (SI Tables~S2 and~S4).

The retained training set comprises \num{\gnNPolyRetained} polymorphs spanning \num{\gnNCompRetained} unique compositions.
Enumerated with the \textsc{pymatgen} vacancy generator on the relaxed unperturbed cell of each, these polymorphs present \num{\gnNSitesClean} symmetry-inequivalent oxygen sites, at each of which a vacancy formation energy was computed.
The sites span twofold through eightfold oxygen coordination (SI Figure~S2), in contrast to the fixed oxygen environments of the perovskite lattice, so the bond-dissociation descriptor defined below varies from site to site through the coordination number.
The retained polymorphs that have ICSD entries outside the near-ambient window are discussed in the Supporting Information.

\subsection{Density Functional Theory Calculations}

All density functional theory (DFT) calculations were performed with the Vienna \textit{Ab~initio} Simulation Package (\textsc{vasp}, version~6.4)\cite{kresse_efficient_1996} using the projector augmented-wave (PAW) method\cite{blochl_projector_1994, kresse_ultrasoft_1999} and the \rtwoscan{} meta-generalized-gradient-approximation (meta-GGA) functional,\cite{furness_accurate_2020,furness_correction_2020} adopted for consistency with the \rtwoscan{} thermodynamic data of the Materials Project.\cite{jain_commentary_2013,kingsbury_performance_2022,kingsbury_flexible_2022}
Input files were prepared with \textsc{pymatgen}\cite{ong_python_2013} following the Materials Project \rtwoscan{} relaxation set,\cite{kingsbury_performance_2022} which applies neither a Hubbard-\(U\) correction nor hybrid exchange.
The plane-wave basis was truncated at a kinetic-energy cutoff of \qty{520}{\eV}, no symmetry was imposed in any calculation, the oxide calculations were spin-polarized, and geometries were optimized until all interatomic forces fell below \qty{0.03}{\eV\per\angstrom}.
The pseudopotentials, the remaining electronic settings, the Brillouin-zone sampling of the bulk cells, and the elemental-metal reference calculations are given in the Supporting Information.
SI Figure~S1 summarizes the full workflow from source structure to fitted model.

Each pristine polymorph was relaxed, both ions and cell, in two independent runs started from the same Materials Project structure: the first from that structure unperturbed, and the second from a perturbed copy, in which both the atomic positions and the lattice vectors were displaced by Gaussian noise with a standard deviation equal to \qty{10}{\percent} of the minimum interatomic distance.
The forces on a symmetric configuration cannot break its symmetry, so only this ``rattle'' run can reach a lower-energy structure of different symmetry. It found one for \gnNSymChangeWord{} of the \num{\gnNPolyTotal} polymorphs, all of them excluded from training (SI Table~S2).
One of the two relaxed cells hosts the vacancy calculations of each retained polymorph and represents that polymorph everywhere below (cohesive energy, reduction potential, and convex hull); the chosen cell is the lower-energy of the two or lies within \qty{1}{\meV\per\atom} of it (SI Tables~S1 and~S2).

\subsection{Vacancy Supercells, Configuration Search, and Formation Energies}

Neutral oxygen vacancies were generated at every symmetry-inequivalent oxygen site of each relaxed polymorph with the \textsc{pymatgen} defect tools.\cite{ong_python_2013,shen_simulating_2024}
Each vacancy was modeled by removing a single oxygen atom from a supercell chosen as the most nearly cubic expansion of the relaxed cell, with \num{\gnNAtomsMin} to \num{\gnNAtomsMax} atoms (SI Table~S1), and the atomic positions were relaxed at fixed cell shape and volume.

To prevent the defect from becoming trapped in a metastable local minimum, the local configuration space around each vacancy was explored with \textsc{ShakeNBreak} using its default protocol.\cite{mosquera-lois_shakenbreak_2022, mosquera-lois_identifying_2023} The bonds between the vacancy and its two nearest cation neighbors were distorted by \qtyrange{-60}{60}{\percent} of their equilibrium length in \qty{10}{\percent} increments (\num{13} levels per site) with a superimposed random rattle of every atom in the supercell, and each resulting configuration was relaxed independently, with the Brillouin zone sampled at the \(\Gamma\) point only; the sites that lack the energy of a level are identified in the Supporting Information.
The lowest-energy relaxed configuration was retained as the ground-state vacancy structure.

This ground-state configuration and a pristine supercell of identical size and shape were then re-relaxed at fixed cell on a \(\Gamma\)-centered mesh, the production mesh (\texttt{KSPACING} = \qty{0.22}{\per\angstrom}, the default spacing of the Materials Project \rtwoscan{} input set). All formation energies entering the fit are evaluated from this dense-matched pair of calculations.
The pristine supercell of every polymorph in the fit was also re-relaxed at fixed cell from a rattled start, to test for a lower-energy configuration that a supercell tiled from the relaxed primitive cell cannot reach; none of the \num{\gnNRefcheckHosts} references relaxed to a lower-energy structure (SI Table~S3).

The neutral oxygen vacancy formation energy was evaluated as
\begin{equation}
  E_\mathrm{v} = E_\mathrm{def} - E_\mathrm{bulk} + \mu_\mathrm{O},
  \label{eq:ev}
\end{equation}
where \(E_\mathrm{def}\) is the total energy of the relaxed defective supercell and \(E_\mathrm{bulk}\) is the total energy of a defect-free supercell of identical size, shape, and $k$-point sampling.\cite{freysoldt_first-principles_2014}
The oxygen chemical potential \(\mu_\mathrm{O}\) is taken at the oxygen-rich limit, as one-half the total energy of an isolated, spin-polarized \ce{O2} molecule computed at the same level of theory, $\mu_\mathrm{O} = \tfrac{1}{2}E(\ce{O2}) = \qty{-5.930}{\eV}$, with no empirical \ce{O2} correction.
Because \rtwoscan{} slightly overbinds \ce{O2}, omitting a correction shifts all \emph{absolute} formation energies by a uniform $\sim$\qty{0.11}{\eV} per oxygen; this constant is absorbed into the model intercept and affects neither the relative trends nor the fitted descriptor coefficients.
No finite-size correction was applied to these neutral defects.\cite{freysoldt_first-principles_2014}

\subsection{Crystal Feature Model: Descriptor Construction}

The neutral oxygen vacancy formation energy is predicted with the crystal feature model of Wexler et al.,\cite{wexler_factors_2021} an interpretable linear regression whose four descriptors are the successive terms of the extended Born--Haber cycle introduced above: breaking the metal--oxygen bonds that coordinate the vacant site, accommodating the two liberated electrons by reducing neighboring cations or by populating the conduction band, and modulating both by the (in)stability of the host oxide.
Two descriptors are \emph{local} (evaluated from the immediate coordination environment of the vacant oxygen), and two are \emph{global} properties of the host crystal.

The \emph{crystal bond dissociation energy sum}, \(\Sigma E_\mathrm{b}\), quantifies the cost of severing the O--M bonds at the vacancy,\cite{wexler_factors_2021}
\begin{equation}
  \Sigma E_\mathrm{b} = \mathrm{CN_O}\cdot E_\mathrm{b}[\ce{O}\text{--}\ce{M}], \qquad
  E_\mathrm{b}[\ce{O}\text{--}\ce{M}] = \frac{E_\mathrm{c}[\ce{MO}_{n/2}]}{N_\mathrm{b}},
\end{equation}
where \(E_\mathrm{c}[\ce{MO}_{n/2}]\) is the cohesive energy of the lowest-energy retained polymorph of the binary oxide \(\ce{MO}_{n/2}\), \(N_\mathrm{b}\) is the number of O--M bonds per formula unit, and the oxygen coordination number \(\mathrm{CN_O}\) is the number of first-nearest-neighbor cations of the vacant oxygen, obtained with the CrystalNN algorithm.\cite{pan_benchmarking_2021}
The cohesive energy was computed by two independent routes: an \rtwoscan-consistent route, from spin-polarized isolated atoms (SI Table~S9), and an experimental route, from tabulated \qty{298}{\kelvin} thermochemistry\cite{barin_thermochemical_1995,chase_nist-janaf_1998} (Supporting Information).

The \emph{crystal reduction potential}, \(V_\mathrm{r}\), quantifies how readily the host's cations accept the two liberated electrons.
For the reduction \(\ce{M}^{n+}\!\rightarrow\ce{M}^{m+}\) (\(n>m\)), referenced to \(\tfrac{1}{4}\)\,\ce{O2}(g), the gas-phase reservoir that also receives the vacancy oxygen in \cref{eq:ev},
\begin{equation}
  V_\mathrm{r} = \frac{-E_\mathrm{r}}{(n-m)\,F}, \qquad
  E_\mathrm{r} = E[\ce{MO}_{m/2}] + \frac{n-m}{4}\,E(\ce{O2}) - E[\ce{MO}_{n/2}],
\end{equation}
with \(F\) the Faraday constant and each oxide energy that of its lowest-energy retained polymorph.\cite{wexler_factors_2021}
Each composition is assigned a single couple, its adjacent recorded one: \(n\) is the average cation oxidation state of the host composition, and \(m\) is the next-lower oxidation state among the single-valence binary oxides of the same metal in the dataset (e.g., \ce{PbO2}\(\rightarrow\)\ce{PbO}), or \(m = 0\), reduction to the elemental metal, when no lower oxide is recorded.

In the perovskite parent, \(V_\mathrm{r}\) is the maximum over the cations neighboring the vacant oxygen;\cite{wexler_factors_2021} in a single-cation oxide, every neighbor is the same metal, so that maximum reduces to one value per composition, and site-to-site variation enters the descriptor set only through the coordination number in \(\Sigma E_\mathrm{b}\).
As with \(E_\mathrm{b}\), \(V_\mathrm{r}\) was evaluated on both the \rtwoscan{} and experimental energy scales; the two global descriptors below are \rtwoscan{} quantities in both routes.

The \emph{band gap} \(E_\mathrm{g}\) captures the cost of promoting electrons into the conduction band when they cannot be fully accommodated by cation reduction.
We used the \rtwoscan{} fundamental gap, converged for every polymorph in the fit with a static calculation on a dense \(\Gamma\)-centered mesh (\(k\)-point spacing \qty{0.11}{\per\angstrom}).
Three hosts, \(Ia\bar{3}\) \ce{Tl2O3}, rutile (\(P4_2/mnm\)) \ce{PbO2}, and \(Pbcn\) \ce{PbO2}, are metallic within \rtwoscan{} and were assigned \(E_\mathrm{g}=0\).

The \emph{energy above the convex hull} \(E_\mathrm{hull}\) (in \unit{\eV\per\atom}) encodes the thermodynamic (in)stability of the host, a modulating factor first incorporated in the perovskite CFM.\cite{wexler_factors_2021}
It is computed at the \rtwoscan{} level, consistent with the vacancy energetics, from our own relaxed retained polymorphs together with the elemental-metal and \ce{O2} references.
The current Materials Project hull energies\cite{jain_commentary_2013} agree with these values to within \qty{\gnEhullMPMaxDiffRT}{\meV\per\atom} for the room-temperature-stable hosts (SI Figure~S12), and substituting them for \(E_\mathrm{hull}\) leaves the cross-validated accuracy of both model variants unchanged, so database hulls can substitute for the descriptor when the model is applied to compositions outside our calculation set.

\subsection{Crystal Feature Model: Regression and Validation}

Mirroring the perovskite analysis,\cite{wexler_factors_2021} two model variants are reported: an all-site model trained on all \num{\gnNSitesClean} sites, which we refer to as the clean sites, and a room-temperature-stable model restricted to the \num{\gnNSitesRT} sites whose host lies within \qty{25}{\meV\per\atom} (\(\approx k_\mathrm{B}T\) at \qty{298}{\kelvin}) of the convex hull.
The four descriptors are combined in a linear model,
\begin{equation}
  E_\mathrm{v} = c_\mathrm{b}\,\Sigma E_\mathrm{b} + c_\mathrm{r}\,V_\mathrm{r} + c_\mathrm{g}\,E_\mathrm{g} + c_\mathrm{hull}\,E_\mathrm{hull} + c_0,
  \label{eq:cfm}
\end{equation}
fit with the Huber regressor,\cite{huber_robust_1964} a robust estimator whose loss is linear for large residuals so that outlying sites cannot dominate the fit, as implemented in \textsc{scikit-learn}.\cite{pedregosa_scikit-learn_2011}
Following the perovskite study,\cite{wexler_factors_2021} the model was cross-validated by averaging over \num{1000} independent fits, each trained on a random \qty{50}{\percent} partition of the sites and evaluated on the held-out half; we report the mean and standard deviation of the held-out (test) mean absolute error (MAE) and of the coefficients, and models are compared on paired splits (the same \num{1000} partitions), by the mean change of the test MAE and the share of splits in which it decreases.
On the present data, the bond-dissociation and hull coefficients of \cref{eq:cfm} are statistically null (see Results), so the primary model retains only \(V_\mathrm{r}\) and \(E_\mathrm{g}\); the full four-descriptor form is reported alongside it, as the Born--Haber decomposition and for comparability with the perovskite parent.

Because three of the four descriptors take a single value per host, a random site-level partition places sites of the same host on both sides of the split. To confirm that this shared-host information does not inflate the held-out accuracy, leave-one-polymorph-out and leave-one-composition-out cross-validation was also performed, in which every site of a held-out group is predicted by a model trained without that group.
Pooled over all held-out sites, the resulting MAEs of the primary model, \qty{\gnLopoRTPooled}{\eV} (leave-one-polymorph-out) and \qty{\gnLocoRTPooled}{\eV} (leave-one-composition-out) for the room-temperature-stable model (\num{\gnLopoAllPooled} and \qty{\gnLocoAllPooled}{\eV} for the all-site model), reproduce the random-split values, so the reported accuracy also describes prediction for compositions absent from the training set (SI Table~S5).

\section{Results and Discussion}

\subsection{Periodic Trends in Vacancy Formation Energies}

\begin{figure*}
  \centering
  \includegraphics[width=\linewidth]{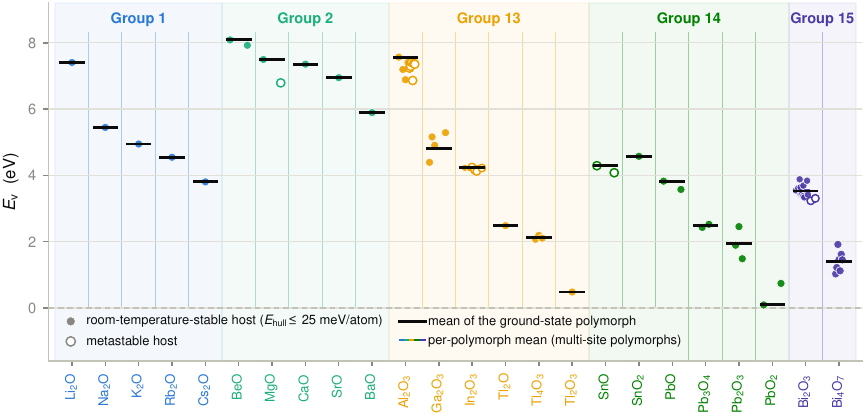}
  \caption{Neutral oxygen vacancy formation energies \(E_\mathrm{v}\) across the binary main-group metal oxides. Each point is one symmetry-inequivalent oxygen site (\num{\gnNSitesClean} clean sites across \num{\gnNPolyClean} polymorphs of \num{\gnNCompRetained} compositions). Compositions are blocked by main group and ordered within each group by cation atomic number, with same-cation stoichiometries ordered by increasing oxygen-to-metal ratio (increasing formal cation oxidation state). Within each composition, sites belonging to the same polymorph are clustered about a common horizontal offset, with polymorphs ordered left to right by increasing energy above the convex hull \(E_\mathrm{hull}\) (the leftmost polymorph is the most stable); horizontal position otherwise carries no quantitative meaning. Short colored ticks give per-polymorph mean \(E_\mathrm{v}\) for multi-site polymorphs in multi-polymorph compositions; the long black bar marks the mean \(E_\mathrm{v}\) of each composition's lowest-\(E_\mathrm{hull}\) (ground-state) polymorph, the value compared across compositions; for \gnMetaGroundStateComps{}, whose polymorphs all lie outside the room-temperature window, that polymorph is drawn with open symbols. The full polymorph list, with the number of sites per polymorph, is given in SI Table~S1.}
  \label{fig:evtrend}
\end{figure*}

The curated dataset yields \num{\gnNSitesClean} well-defined neutral oxygen vacancy formation energies across \num{\gnNCompRetained} binary main-group oxide compositions.
\Cref{fig:evtrend} organizes these energies by main group and composition, revealing trends down each main group and with cation oxidation state.
The periodic trend is read from each composition's lowest-\(E_\mathrm{hull}\) (ground-state) polymorph, the thermodynamically preferred phase, rather than from a coverage-weighted average over every computed site.
The formation energies span \qty{\gnEvSpan}{\eV}, from \qty{0.09}{\eV} for a vacancy in the rutile (\(P4_2/mnm\)) polymorph of \ce{PbO2} to \qty{8.09}{\eV} in wurtzite \ce{BeO}.
Within each main group, the \(E_\mathrm{v}\) of the ground-state polymorph decreases monotonically down the group as the cation grows and its charge density drops: from \ce{Li2O} (\qty{7.4}{\eV}) to \ce{Cs2O} (\qty{3.8}{\eV}) across group 1, and from \ce{BeO} (\qty{8.1}{\eV}) to \ce{BaO} (\qty{5.9}{\eV}) across group 2.
Among the multivalent \ce{Tl}, \ce{Pb}, and \ce{Bi} oxides, the ground-state-polymorph \(E_\mathrm{v}\) also decreases monotonically with increasing cation oxidation state; \ce{Sn} is the exception, with its two oxidation states within \qty{\gnSnOtwoMinusSnO}{\eV} of each other and the dioxide slightly higher.
Metastable polymorphs of the same composition lie mostly below these ground-state values (open symbols in \cref{fig:evtrend}) without disturbing the trend.
The closed-shell, redox-inert oxides of groups 1 and 2 are the most vacancy-resistant (means over all sites of \qty{\gnEvMeanGroupOne}{\eV} for group 1 and \qty{\gnEvMeanGroupTwo}{\eV} for group 2), whereas the reducible late-\(p\)-block oxides of groups 13--15 form vacancies far more readily; the group-14 (\ce{Pb}, \ce{Sn}) and group-15 (\ce{Bi}) oxides are the lowest (all-site means of \qty{\gnEvMeanGroupFourteen}{\eV} and \qty{\gnEvMeanGroupFifteen}{\eV}, respectively), with the lowest sites (both \ce{PbO2} polymorphs and \ce{Tl2O3}) falling below \qty{1}{\eV}.

The scatter among the sites of a single polymorph in \cref{fig:evtrend} also reflects chemical differences among the groups.
The group 1 and 2 oxides show no such scatter, because the isotropic, ionic bonding of these closed-shell cations favors high-symmetry structures (antifluorite, anti-\ce{CdCl2}, rock salt, wurtzite, and rutile-related) with a single symmetry-inequivalent oxygen site.
The group 13 sesquioxides, whose smaller trivalent cations bond more covalently and directionally, instead adopt polymorph-rich, lower-symmetry structures whose inequivalent oxygen sites differ in local environment (within \ce{Al2O3} alone, oxygen coordination numbers of \gnAlOCNmin{} to \gnAlOCNmax{} and bond-dissociation sums \(\Sigma E_\mathrm{b}\) spanning \qtyrange[range-phrase = --, range-units = single]{\gnAlOSigmaEbMin}{\gnAlOSigmaEbMax}{\eV}), spreading \(E_\mathrm{v}\) by as much as \qty{\gnGaOSpread}{\eV} within a single composition (\ce{Ga2O3}).
The site diversity that produces this scatter is the local-environment variation the crystal feature model's local descriptors are constructed to capture; within a single host, it is also the variation that they ultimately fail to capture (see the outlier analysis below).

The organizing variable is cation reducibility.
Oxides whose metal can access a lower oxidation state (\ce{Pb^{4+}}\(\rightarrow\)\ce{Pb^{2+}}, \ce{Tl^{3+}}\(\rightarrow\)\ce{Tl^{+}}) form vacancies far more easily than oxides of the redox-inert closed-shell cations of groups 1 and 2, because the two electrons left behind by the departing oxygen can occupy an energetically accessible state.
This is the qualitative signature that the crystal reduction potential \(V_\mathrm{r}\) is constructed to quantify, and it motivates the descriptor analysis that follows.

\subsection{The Crystal Feature Model}

In isolation, three of the four descriptors correlate with \(E_\mathrm{v}\) with the signs that the extended Born--Haber cycle predicts (SI Figure~S10).
The crystal reduction potential is the strongest single predictor (Pearson \(R = \gnRVrDFT\)), reproducing its dominance in the perovskite CFM (\(R = -0.88\)).\cite{wexler_factors_2021} The band gap is second and positive (\(R = \gnREgDFT\)), far stronger than in the perovskite study (\(R = +0.34\)),\cite{wexler_factors_2021} because the main-group binaries span band gaps from zero (the three metallic hosts) to \qty{\gnEgMax}{\eV}, a far wider range than that of the transition-metal perovskites. The bond-dissociation sum correlates moderately (\(R = \gnREbDFT\)), whereas the energy above the hull correlates only weakly and with the opposite sign to the one the cycle predicts (\(R = \gnREhullDFT\)).

\begin{figure}
  \centering
  \includegraphics[width=\linewidth,height=0.72\textheight,keepaspectratio]{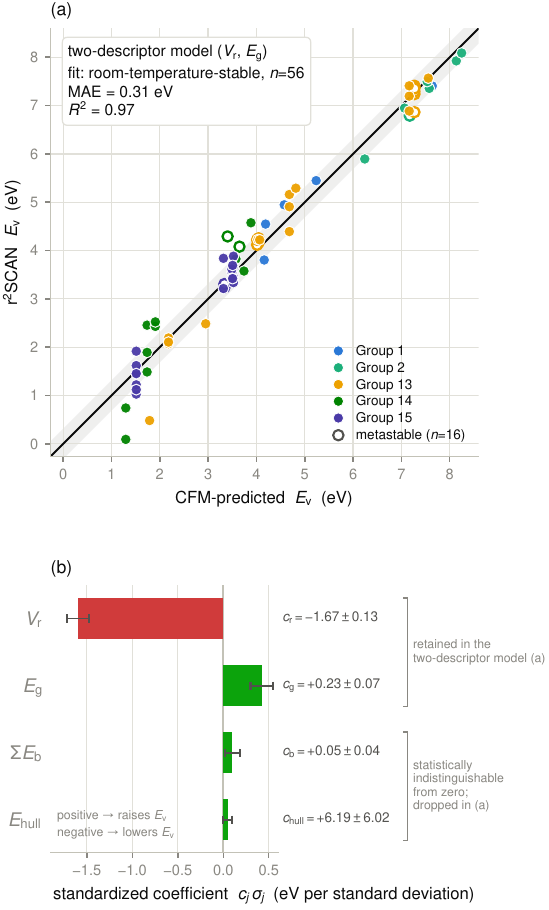}
  \caption{The binary-oxide crystal feature model (CFM). (a) \rtwoscan{} versus predicted \(E_\mathrm{v}\) for the primary two-descriptor model of \cref{eq:cfm-fitted}, colored by main group. Filled points are the \num{\gnNSitesRT} room-temperature-stable sites on which the primary model is fit (held-out test MAE \qty{\gnMaePrimary}{\eV}, \(R^2 = \gnRTwoPrimRT\)); open points are the \num{\gnNSitesMeta} additional metastable sites, which lie outside the fit, so their predictions are out of sample (MAE \qty{\gnMaeMetastable}{\eV}). The band marks \(\pm\)MAE about the parity line. (b) Standardized coefficients \(c_j\sigma_j\) of the full four-descriptor room-temperature-stable fit, each the change in \(E_\mathrm{v}\) per one standard deviation \(\sigma_j\) of its descriptor (\qty{\gnSigmaVr}{\volt} for the reduction potential \(V_\mathrm{r}\), \qty{\gnSigmaEg}{\eV} for the band gap \(E_\mathrm{g}\), \qty{\gnSigmaEhull}{\meV\per\atom} for the hull energy \(E_\mathrm{hull}\), and \qty{\gnSigmaEb}{\eV} for the bond-dissociation sum \(\Sigma E_\mathrm{b}\)); the native-unit coefficients \(c_j\) and their fit uncertainties are annotated at the right.}
  \label{fig:parity}
\end{figure}

Combining the four descriptors in the linear model of \cref{eq:cfm} reproduces the \rtwoscan{} formation energies, but two of the terms carry no statistical weight.
For the room-temperature-stable hosts, the four-descriptor fit gives a held-out test MAE of \qty{\gnMaeFourRTUnc}{\eV} (mean over the \num{1000} splits; standard deviation in the last digit; \(n = \num{\gnNSitesRT}\)), with bond-dissociation and hull coefficients both within two standard deviations of zero (the consensus fit to all those sites, an in-sample measure, has \(R^2 = \gnRTwoFourRT\)). Trained on all \num{\gnNSitesClean} clean sites, metastable hosts included, it gives \qty{\gnMaeFourAllUnc}{\eV} (in-sample \(R^2 = \gnRTwoFourAll\)) with the same two null terms.
Restricting the model to its two surviving descriptors lowers the held-out error: on paired splits the test MAE changes by \qty{\gnDropBothRTMean}{\eV} for the room-temperature-stable sites (lower in \qty{\gnDropBothRTPctLower}{\percent} of the splits) and by \qty{\gnDropBothAllMean}{\eV} over all sites (\qty{\gnDropBothAllPctLower}{\percent}).
The primary model, fit to the room-temperature-stable sites, is
\begin{equation}
  E_\mathrm{v} = \gnPrimCr\,V_\mathrm{r} + \gnPrimCg\,E_\mathrm{g} + \gnPrimCzero,
  \label{eq:cfm-fitted}
\end{equation}
with \(E_\mathrm{v}\) and \(E_\mathrm{g}\) in \unit{\eV} and \(V_\mathrm{r}\) in \unit{\volt}; it achieves a test MAE of \qty{\gnMaePrimRTUnc}{\eV} on those sites (in-sample \(R^2 = \gnRTwoPrimRT\); \cref{fig:parity}a); refit on all \num{\gnNSitesClean} clean sites, the same two-descriptor form gives \qty{\gnMaePrimAllUnc}{\eV} (in-sample \(R^2 = \gnRTwoPrimAll\)).
The accuracies of the two training windows are statistically indistinguishable.
The reported accuracy therefore does not depend on the stability restriction.
Both values are lower than those of the four-descriptor transition-metal perovskite parent, which achieved \qty{\gnParentMaeRT(\gnParentMaeRTStd)}{\eV} at the SCAN\,+\,\(U\) level for its room-temperature-stable set and \qty{\gnParentMaeAll(\gnParentMaeAllStd)}{\eV} over all perovskites,\cite{wexler_factors_2021} although the two settings differ: the present dataset contains no transition metals, and the model has no empirical Hubbard correction.

The coefficients of the full four-descriptor fit recover the physical hierarchy of the extended Born--Haber cycle and justify the two-descriptor simplification (\cref{fig:parity}b).
The reduction potential carries by far the largest standardized weight (\qty{\gnStdWeightVr}{\eV} per standard deviation of \(V_\mathrm{r}\), \(\sigma_\mathrm{r} = \qty{\gnSigmaVr}{\volt}\) over the \num{\gnNSitesRT} room-temperature-stable sites; native \(c_\mathrm{r} = \gnCrNative\) in the room-temperature-stable fit), confirming that electron accommodation by cation reduction dominates the energetics, and its native coefficient has the same sign as the perovskite value (\(\approx \gnParentCr\)) and a magnitude about \num{\gnCrMinusParent} larger.\cite{wexler_factors_2021}
The band gap is the second most important term (\(c_\mathrm{g} = \gnCgRT\)), on a meta-GGA scale of the same order as the SCAN\,+\,\(U\)-level \(c_\mathrm{g} \approx 0.4\) of the parent model.

The bond-dissociation term is null (\(c_\mathrm{b} = \gnCbRT \pm \gnCbRTStd\)): although \(\Sigma E_\mathrm{b}\) correlates moderately with \(E_\mathrm{v}\) on its own, \(V_\mathrm{r}\) and \(E_\mathrm{g}\) already capture its variance, and removing it changes the room-temperature-stable test MAE by \qty{\gnDropEbRTMean}{\eV} on paired splits (lower in \qty{\gnDropEbRTPctLower}{\percent} of them). This parallels the perovskite study, whose simplified room-temperature-stable model (built for the case of unknown \(E_\mathrm{g}\) and retaining only \(V_\mathrm{r}\) and \(E_\mathrm{hull}\)) dropped the bond term because its coefficient, \(-0.03 \pm 0.15\), was statistically indistinguishable from zero.
The hull coefficient is also null: its mean lies within two standard deviations of zero in both the room-temperature-stable fit (\(\gnChullRT \pm \gnChullRTStd\)) and the all-site fit (\(\gnChullAll \pm \gnChullAllStd\)), and it is as roster-dependent here as in the perovskite model, where it ranged from \(\gnParentChullAll\) to \(\gnParentChullRT\).\cite{wexler_factors_2021}
This instability reflects redundancy rather than missing physics: a signed counterpart of \(E_\mathrm{hull}\), the per-oxygen energy of reduction to the adjacent oxygen-poorer phases on our \rtwoscan{} convex hull, is collinear with \(V_\mathrm{r}\) (\(R = \gnRRedVrAll\) over all \num{\gnNSitesClean} clean sites; Supporting Information).
\(E_\mathrm{hull}\) therefore contributes only the polymorph-level metastability that \(V_\mathrm{r}\) does not already encode; within the room-temperature window, that remainder spans \qtyrange[range-phrase = --, range-units = single]{\gnEhullRTmin}{\gnEhullRTmax}{\meV\per\atom}, too narrow a range to constrain a coefficient.

Dropping both null terms yields the primary model of \cref{eq:cfm-fitted}; restricting the fit to \(V_\mathrm{r}\) alone, by contrast, degrades the test MAE to \qty{\gnMaeVrOnlyRTUnc}{\eV} (a paired change of \qty{\gnDropEgRTMean}{\eV}, lower in only \qty{\gnDropEgRTPctLower}{\percent} of the splits), so the band gap contributes independent predictive information.
Finally, models built from the experimental-thermochemistry route, which changes only \(\Sigma E_\mathrm{b}\) and \(V_\mathrm{r}\), have test MAEs \qtyrange[range-phrase = { to }, range-units = single]{\gnExpRouteDiffMin}{\gnExpRouteDiffMax}{\eV} above those of the \rtwoscan{} route in the four fits (two- and four-descriptor, both windows), so \(V_\mathrm{r}\) can be sourced from tabulated thermochemistry at a modest cost in accuracy; the formation enthalpies of the mixed-valence \ce{Pb2O3}, \ce{Tl4O3}, and \ce{Bi4O7}, which carry \num{\gnNSitesEstimatedDHf} of the sites, are estimates rather than tabulated values.

The reported accuracy was also verified to be robust to the methodological choices that most affect the target energies or the band-gap descriptor: the pristine reference, the band-gap definition, and the production \(k\)-point sampling (SI Figure~S11).
Evaluated on the four-descriptor form and on the same paired splits, the test MAE moves from \qty{\gnLadderARTUnc}{\eV} with a \(\Gamma\)-only perfect-supercell reference to \qty{\gnLadderERTUnc}{\eV} with the adopted convention.
Varying the stability window shows the accuracy is not an artifact of the window choice: as the \(E_\mathrm{hull}\) cutoff is relaxed from \qty{25}{\meV\per\atom} through \num{50} and \qty{100}{\meV\per\atom} to no cutoff at all, the within-window test MAE of the primary model takes the values \num{\gnSweepPrimTwentyFive}, \num{\gnSweepPrimFifty}, \num{\gnSweepPrimHundred}, and \qty{\gnSweepPrimAll}{\eV} (Supporting Information).
We retain the \qty{25}{\meV\per\atom} window for comparability with the parent model.

\subsection{Outliers, Within-Host Trends, and the Limits of a Linear Model}

An interpretable model's residuals can be read mechanistically. Those of this model separate into three classes, each examined as a distinct route to extending the model: computational artifacts, which the residuals themselves exposed and curation removed or corrected; a systematic limit on how finely the model distinguishes sites within the same host, probed by searching for a fifth, site-resolved descriptor; and genuine electron-accommodation physics beyond the descriptor set, failures in \emph{both} directions that together identify the missing term.

\subsubsection{Computational Artifacts in the Residuals}

Two vacancy calculations exposed unstable structural models of their hosts; in a fit that includes them, their three sites carry the largest residuals of the dataset.
In the fluorite-type (\(Fm\bar{3}m\)) cell of \ce{PbO2} and in the orthorhombic (\(Cmc2_1\)) cell of red \ce{SnO}, the vacancy triggered a global reconstruction of the entire host rather than any local rebonding (SI Figures~S8 and~S9), and both cells are themselves unstable in our relaxations.
Both polymorphs were therefore excluded during curation; their collapses are a caution that a vacancy energy computed in an incorrect or unstable structural model of the host reflects the relaxation of the host away from its assumed cell rather than defect chemistry.
The \(Pbca\) cell of red \ce{SnO} from the single-crystal redetermination, computed with the identical workflow and retained in the training set, shows none of this pathology.

\subsubsection{Within-Host Degeneracy and the Search for a Site-Resolved Descriptor}

The second class is subtler and systematic: the two-descriptor model is exactly flat \emph{within} each polymorph, because both of its descriptors are host-level quantities (in the four-descriptor form, site-to-site discrimination appears only through \(\Sigma E_\mathrm{b}\), whose null coefficient suppresses it).
The observed within-host spreads are substantial: the \gnNSitesBiSevenWord{} sites of \ce{Bi4O7} span \qty{\gnSpanBiSeven}{\eV}, the \gnNSitesGaMonoWord{} sites of monoclinic \ce{Ga2O3} span \qty{\gnSpanGaMono}{\eV}, and the \gnNSitesPbMonoWord{} sites of the mixed-valence \ce{Pb2O3} span \qty{\gnSpanPbMono}{\eV}, each set at a single predicted value, so this descriptor degeneracy sets a floor on the within-host resolution of any model built from these quantities.
We therefore searched systematically for a fifth, site-resolved descriptor (SI Figure~S13).

Each candidate was judged on host-centered values: the mean over a polymorph's sites is subtracted from both the descriptor and \(E_\mathrm{v}\), leaving only site-to-site differences (\num{\gnNCenPts} points from the \num{\gnNCenHosts} polymorphs with at least two sites).
A single global coefficient can describe a within-host trend only if that trend has the same sign in every host, so each candidate's correlation was also counted host by host, over the hosts with at least three sites (the \(+a/-b\) annotations of SI Figure~S13a).
The candidates are initial-state properties of the pristine crystal (the reduction potential of the vacancy's cation shell, coordination numbers and bond sums, bond lengths and bond-valence sums, the Madelung potential, the geometry of the cation shell, and the O-2p band center) and final-state properties of the relaxed defect supercell (the Kohn--Sham donor depth, the localization of the defect level, and the relaxation displacements); their definitions are given in the Supporting Information.
Most candidates fail not because their correlations are weak but because their signs are inconsistent across hosts, so a global coefficient fit to one subset of hosts mispredicts the rest.

Among the sign-consistent candidates, the shortest metal--oxygen bond length of the vacant site, \(d_\mathrm{min}\), is the strongest: its host-centered \(R = \gnDminR\) stays between \(\gnDminRLooMin\) and \(\gnDminRLooMax\) when any one host is left out, and its correlation is positive in each of the \gnNHostsThreeWord{} hosts with at least three sites (SI Figure~S13).
Within a host, the overbonded oxygen sites, those with the shortest bonds, are systematically the cheaper vacancies, a local strain-relief signature that runs opposite to the bond-breaking expectation encoded in \(\Sigma E_\mathrm{b}\).
The bond length itself, not a proxy, carries the signal: within hosts, with the integer \(\mathrm{CN_O}\), \(R(E_\mathrm{v}, d_\mathrm{min}\,|\,\mathrm{CN_O}) = \gnPartialDminGivenCN\) whereas \(R(E_\mathrm{v}, \mathrm{CN_O}\,|\,d_\mathrm{min}) = \gnPartialCNGivenDmin\), so the coordination number is a proxy for the bond length, not the reverse.
This ordering does not, however, translate into accuracy: adding \(d_\mathrm{min}\) to the primary model changes its held-out test MAE on the room-temperature-stable sites by only \qty{\gnDminAddDMae}{\eV}, because the magnitude of the within-host trend differs from host to host (Supporting Information).

A site-resolved reduction potential makes the model worse.
Replacing the composition-level \(V_\mathrm{r}\) by an oxidation-state-resolved average over the vacancy's first cation shell (which differs from the composition value only in the mixed-valence hosts) \emph{degrades} the four-descriptor model (test MAE \qty{\gnSiteVrMaeRTBase}{\eV} \(\rightarrow\) \qty{\gnSiteVrMaeRT}{\eV} on the room-temperature-stable sites).
The local and global trends need not share a sign: across compositions, a more reducible cation lowers \(E_\mathrm{v}\), but in \ce{Pb2O3}, where \ce{Pb^{4+}} is by far the more reducible of the two cations, vacancies coordinated by more \ce{Pb^{4+}} are \emph{costlier} to form (SI Figure~S7), because the stronger binding of oxygen to its higher-charged neighbors outweighs the greater reducibility of the shell.
The host-level \(V_\mathrm{r}\) is thus the appropriate formulation: the adjacent reduction couple available to the two released electrons is a property of the crystal as a whole, not an average over the vacancy's first coordination shell.

Outside the bond-geometry family, the Kohn--Sham donor depth is the only sign-consistent candidate, but it overlaps with \(d_\mathrm{min}\): in leave-one-host-out tests, the two together recover \qty{\gnLohoDminGapAll}{\percent} of the host-centered variance, against \qty{\gnLohoDminAll}{\percent} for \(d_\mathrm{min}\) alone.
No single descriptor from a published library\cite{ward_matminer_2018} passes the false-discovery-rate, sign-consistency, and leave-one-host-out tests (Supporting Information).

The final-state electronic structure limits what any site-resolved descriptor can recover, which is why the search saturates, and at the same time confirms the Born--Haber picture the model is built on.
The neutral vacancy forms no polaron at any site, and its defect level is a single doubly occupied band at \num{\gnNSingleBand} of the \num{\gnNSitesClean} sites, deep in the gap at all but \gnNShallowWord{} sites of the gapped hosts.
At \num{\gnNSLike} of the \num{\gnNSitesClean} sites, the metal component of the level is predominantly \(s\)-like (more than half of the metal weight), the cation-reduction step of the cycle made directly visible.
The site-to-site ordering of \(E_\mathrm{v}\) thus tracks the depth of this defect level, which in turn tracks the local bond geometry that \(d_\mathrm{min}\) encodes, but the \emph{magnitude} of that dependence is host-specific.
The remaining within-host variation (a root-mean-square value of \qty{\gnResidRmsRT}{\eV} in the room-temperature window after a pooled \(d_\mathrm{min}\) term fitted to these same sites) is therefore irreducible for any single global linear term and points instead to hierarchical, per-host-calibrated extensions.

\subsubsection{The Electron-Accommodation Boundary}

The third class carries the two largest residuals among the room-temperature-stable sites and identifies the physics the two descriptors approximate most coarsely: electron accommodation.
The three hosts that \rtwoscan{} predicts to be metallic, \(Ia\bar{3}\) \ce{Tl2O3}, rutile (\(P4_2/mnm\)) \ce{PbO2}, and \(Pbcn\) \ce{PbO2}, have \gnMetallicLowestPhrase{} formation energies in the dataset (\num{\gnEvTlIa}, \num{\gnEvPbRut}, and \qty{\gnEvPbPbcn}{\eV}), which the primary model overpredicts by \qtyrange[range-phrase = --, range-units = single]{\gnMetallicOverMin}{\gnMetallicOverMax}{\eV} (\cref{fig:parity}a).
These low energies reflect how the released electrons are accommodated.
In each host, the two electrons released by the vacancy fill the metallic conduction states at the Fermi level rather than localizing on a reduced cation: the \textsc{ShakeNBreak} distortion sets are flat (every relaxed distortion lies within \qty{\gnMetallicFanSpread}{\meV} of the ground state, apart from one \qty{+60}{\percent} \ce{Tl2O3} endpoint that relaxed into a distinct O--O-dimerized configuration \qty{\gnTlOThreeDimerDE}{\eV} higher) and the local magnetic moments are zero.
Independent defect studies corroborate both the low formation energy and the mechanism: hybrid-functional calculations find the oxygen vacancy in \ce{PbO2} and \ce{Tl2O3} to be an always-ionized donor whose level is resonant in the conduction band (with a neutral formation energy of only \qty{0.8}{\eV} in \ce{Tl2O3} even under oxygen-rich conditions),\cite{scanlon_nature_2011,kehoe_nature_2011} and experiment finds \ce{PbO2} intrinsically oxygen-deficient: neutron diffraction resolves \qty{1.6}{\percent} oxygen substoichiometry.\cite{scanlon_nature_2011}
Facile vacancy formation with conduction-band electron accommodation is thus the established chemistry of these hosts; at the hybrid level, rutile \ce{PbO2} (\qty{0.23}{\eV}) and \ce{Tl2O3} (\qty{0.33}{\eV}) are small-gap degenerate semiconductors rather than \rtwoscan{} metals,\cite{scanlon_nature_2011,kehoe_nature_2011} a distinction that shifts \(E_\mathrm{v}\) by tenths of an electronvolt but leaves the accommodation physics unchanged.

The linear model cannot reproduce such low values.
Every reduction potential in the dataset is negative and \(c_\mathrm{r} < 0\), so the reduction term always \emph{raises} the predicted energy; with \(E_\mathrm{g}=0\) for these metallic hosts, the model cannot predict a formation energy below the intercept of \cref{eq:cfm-fitted} (\qty{\gnPrimCzero}{\eV}), and all three computed metallic-host values lie below that floor, a limit built into the model's functional form rather than a failure of the descriptors.

Electron accommodation can also fail in the opposite direction.
Litharge \ce{SnO} (\(P4/nmm\); a metastable host outside the room-temperature window) is the mirror image of the metallic hosts: the primary model \emph{under}predicts it by \qty{\gnSnOLithUnder}{\eV} (predicted \num{\gnSnOLithPred} versus a computed \qty{\gnSnOLithEv}{\eV}), because its reducible \ce{Sn^{2+}} couple and small band gap suggest low-cost electron accommodation that the actual electronic structure does not provide.
The two electrons instead enter a filled, \ce{Sn}-5s-derived state at the top of the valence band rather than executing the \ce{Sn^{2+}}\(\rightarrow\)\ce{Sn^{0}} reduction assumed in \(V_\mathrm{r}\).
An independent hybrid-functional study corroborates both the value and its interpretation: re-referenced to our oxygen chemical potential, the hybrid-functional neutral vacancy formation energy of Allen et al.\cite{allen_understanding_2013} is \(\approx\)\qty{\gnSnOAllenVO}{\eV}, within \qty{\gnSnOAllenDiff}{\eV} of our \rtwoscan{} result, and the same study finds the oxygen vacancy in \ce{SnO} to be a nondoping defect;\cite{allen_understanding_2013} at the hybrid level, its donor levels lie below the valence-band maximum,\cite{allen_understanding_2013} so the neutral vacancy we compute is also the physically relevant charge state at every attainable Fermi level.

The two failure directions thus bracket a single deficiency, which lies in the accommodation route rather than in the band gap itself: \(V_\mathrm{r}\) and \(E_\mathrm{g}\) are host-level proxies for electron accommodation, and the true cost of accommodating the two liberated electrons can be lower than they encode (itinerant states at the Fermi level of a metallic host) or higher (a filled state at the top of the valence band).
An explicit accommodation-energy term is therefore the extension of the descriptor set to which both failure directions point.
Metallicity per se, however, is not the boundary of the linear model's validity: its perovskite parent was fit across hosts ranging from wide-band-gap insulators to metallic systems.\cite{wexler_factors_2021}
The boundary is the accommodation mode: the model remains valid wherever the liberated electrons localize on a cation redox couple that \(V_\mathrm{r}\) describes, and fails where the accommodating state departs from that couple in either direction; conducting hosts whose carriers localize on a redox-active sublattice, as in many transition-metal oxides, should therefore remain within its reach.

\subsection{Comparison with Prior Models}

The CFM's accuracy of \qty{\gnMaePrimary}{\eV} is competitive with that of the random-forest models trained on much larger and more heterogeneous defect datasets (for example, the \qtyrange[range-phrase = --, range-units = single]{0.27}{0.44}{\eV} of Kumagai et al.\cite{kumagai_insights_2021} across \num{937} oxides and the comparable accuracy of Baldassarri et al.\cite{baldassarri_oxygen_2023} over more than \num{2500} vacancy calculations) but the CFM achieves it with two physically transparent descriptors, no defect calculation at prediction time, and residuals that can be read mechanistically.
The Born--Haber form, transferred from transition-metal perovskites to transition-metal-free main-group binaries and re-evaluated at \rtwoscan, retains the coefficient hierarchy of its surviving terms while its bond-breaking and stability terms prove statistically null, so the four-descriptor parent reduces to a two-descriptor model with no loss of accuracy; the persistent dominance of the reduction potential supports the transferability of the underlying physical picture.
This comparison is not like for like: the parent model predicted perovskite vacancies from the thermochemistry of the cations' binary oxides, a transfer between compound classes, whereas here the descriptors are bulk properties of the same binary oxides whose vacancies are predicted, and \(V_\mathrm{r}\) in particular is built from the host's own bulk energy, so a lower error than the parent's is expected.
The scope of the validation is set accordingly: the descriptors contain no defect calculation of any host, and the leave-one-composition-out error of \qty{\gnLocoRTPooled}{\eV} shows that a held-out composition's vacancy energetics are predicted from bulk properties alone.
What the accuracy demonstrates is therefore the bulk-to-defect map within the binary-oxide class, not transfer between compound classes.

The descriptors themselves are not interchangeable with projected quantum-chemical bond-strength measures, which quantify only the covalent contribution to bond dissociation: on the \num{15} training-set hosts also covered by the \textsc{lobster} bonding database of Naik et al.,\cite{naik_quantum_2023} the integrated crystal orbital Hamilton population (ICOHP) summed over the bonds at the oxygen site is uncorrelated with both \(E_\mathrm{v}\) (\(R = +0.02\)) and \(\Sigma E_\mathrm{b}\) (\(R = +0.05\); SI Table~S8).

\section{Conclusions}

This work has surveyed neutral oxygen vacancy formation at the \rtwoscan{} level in \num{\gnNPolyRetained} polymorphs of the binary main-group metal oxides and obtained \num{\gnNSitesClean} formation energies, ordered across the periodic table by the reducibility of the cation.
A regression of these energies on the successive terms of the extended Born--Haber cycle assigns the bond-breaking and stability terms coefficients statistically indistinguishable from zero.
The largest residuals among the room-temperature-stable sites arise where the liberated electrons, rather than reduce a cation, occupy itinerant conduction states in \(Ia\bar{3}\) \ce{Tl2O3}, rutile (\(P4_2/mnm\)) \ce{PbO2}, and \(Pbcn\) \ce{PbO2}, whose formation energies are overpredicted.
Electron accommodation fails in the opposite direction in litharge \ce{SnO}, where a filled, \ce{Sn}-5s-derived state at the top of the valence band leaves the formation energy underpredicted.
The crystal feature model, linear in the crystal reduction potential and the band gap, has therefore proved valid wherever a cation redox couple accepts the liberated electrons.
Within a host, the shortest metal--oxygen bond of the vacant site orders the formation energies in every polymorph with at least three sites, but the magnitude of that dependence is host-specific, so site-resolved prediction calls for per-host calibration rather than a fifth global descriptor.

Because its inputs are properties of the pristine bulk (and the reduction potential can be taken directly from tabulated experimental thermochemistry, which comes within \qty{\gnExpRouteDiffMax}{\eV} of the first-principles fit in test MAE), the model supplies defect thermodynamics for compositions and polymorphs outside its training set at negligible cost, within the electron-accommodation boundary established above.
Beyond the model, the dataset itself (\num{\gnNSitesClean} curated, \textsc{ShakeNBreak}-optimized vacancy formation energies spanning \qty{\gnEvSpan}{\eV} of chemical variation) offers a benchmark for the machine-learning interatomic potentials now being extended to defect properties, where reference data for simple binary systems remain scarce; the model, in turn, provides a physics-based prior against which such potentials can be regularized or \(\Delta\)-learned (trained to correct its residual) in that data-poor regime.
Extension of both the dataset and the model to the transition-metal binary oxides, where multiple accessible oxidation states and localized \(d\) electrons will test the reduction-potential descriptor most severely, is underway and will be reported separately.

\section*{Data Availability Statement}

The machine-readable data files described in the final section of the Supporting Information accompany this article as \texttt{cfm\_si\_data.zip}.
The same tables, the analysis scripts that regenerate every table and figure, the figure files, and the software-environment specifications will be deposited in a Zenodo record on publication, under the CC BY 4.0 license for the data and figures and the MIT license for the scripts.
The \textsc{vasp} input and output files of every production \rtwoscan{} calculation, except the \texttt{POTCAR} and wavefunction (\texttt{WAVECAR}) files, will be deposited in the NOMAD repository as a dataset on publication; the Supporting Information lists the calculations that it comprises.

\section*{Supporting Information}

The Supporting Information is available free of charge.

Supporting Information (PDF): extended computational methods; the polymorph inventory with its ICSD provenance and the evidence for each curation outcome (SI Table~S4); supercell dimensions and $k$-point sampling (SI Table~S1); the space-group reproduction audit and the reference-supercell check (SI Tables~S2 and~S3); single-descriptor correlations and model-robustness tests (SI Figures~S10 and~S11 and~Table~S5); the comparison of \rtwoscan{} and Materials Project hull energies (SI Figure~S12); the candidate site descriptors and their evaluation conventions (SI Tables~S6 and~S7); the comparison with quantum-chemical bonding descriptors (SI Table~S8); and the isolated-atom reference energies (SI Table~S9).

Machine-readable data files (ZIP): \texttt{cfm\_si\_data.zip}, holding the fitted coefficients and cross-validation statistics, the per-site predictions and descriptor values, the source data of the tables and figures, and the relaxed structures with their absolute total energies, each file described in the final section of the Supporting Information.

\section*{Acknowledgments}

The authors thank the members of the Wexler group for helpful discussions.
This research was supported by the U.S.\ Department of Energy, Office of
Energy Efficiency and Renewable Energy, Hydrogen and Fuel Cell Technologies
Office, under Award No.~DE-EE0010733.
A portion of the research was performed using computational resources
sponsored by the Department of Energy's Office of Energy Efficiency and
Renewable Energy and located at the National Renewable Energy Laboratory.
The high-performance computing facilities at Washington University in
St.~Louis provided additional computational resources.

\bibliography{references}

\newpage

\begin{minipage}[b][1.75in]{3.25in}
  \centering
  \includegraphics[width=3.25in,height=1.75in,keepaspectratio]{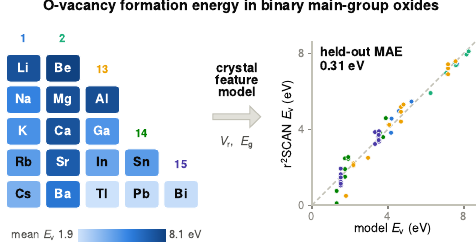}
\end{minipage}

\includepdf[pages=-]{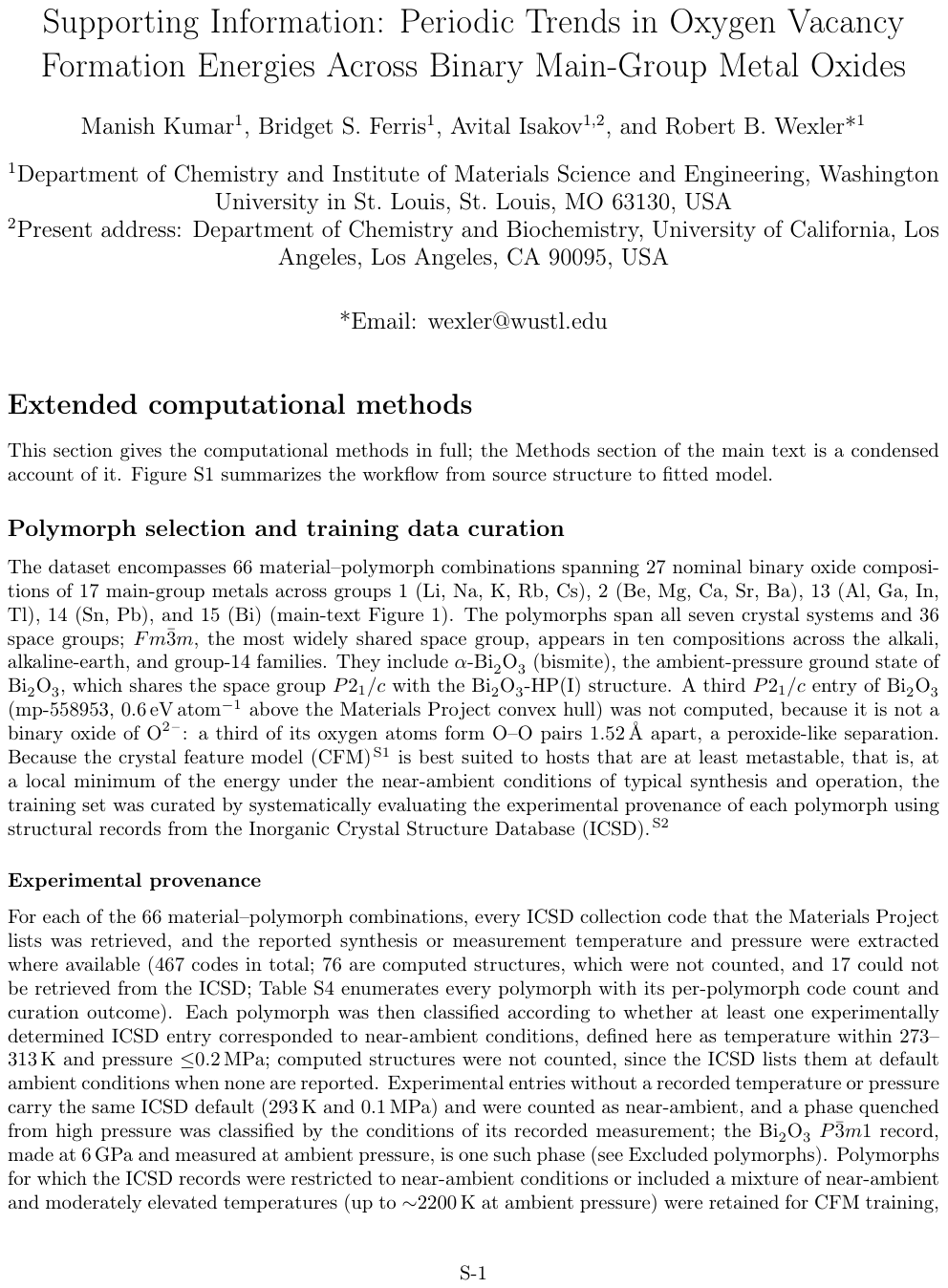}

\end{document}